\documentclass[12pt,a4paper,aps,preprint,superscriptaddress,nofootinbib]{revtex4-1}
\usepackage[utf8]{inputenc}
\usepackage{graphicx}
\usepackage{amssymb}
\usepackage{textcomp}
\usepackage{amsmath}
\usepackage{tabularx}
\usepackage{bm}
\usepackage{times}
\usepackage{color}
\usepackage{ulem}
\usepackage{slashed}
\usepackage{multirow}
\usepackage{verbatim}
\usepackage{cancel}
\usepackage{subfigure}
\usepackage{epstopdf}
\usepackage{mathrsfs}
\usepackage{amsmath}
\usepackage{geometry}
\usepackage{graphicx}
\usepackage[none]{hyphenat}
\usepackage{appendix}
\usepackage{ragged2e}
\usepackage{url} 
\usepackage{tikz}
\usetikzlibrary{shapes,snakes}
\usepackage{booktabs}
\usepackage{parskip}
\usepackage{upgreek}
\usepackage{setspace}

\usepackage{amsmath,physics,booktabs}
\usepackage{array}
\newcolumntype{C}{>{$}c<{$}}

\usepackage[colorlinks=true, pdfstartview=FitV, linkcolor=blue, citecolor=blue, urlcolor=blue]{hyperref}
\allowdisplaybreaks[4]

\newcommand{\ba}{\begin{eqnarray}}
	\newcommand{\ea}{\end{eqnarray}}
\newcommand{\be}{\begin{equation}}
	\newcommand{\ee}{\end{equation}}

\begin{document}	
	\title{Magnetoelectric Phase Transition and Axion Dynamics}
	
	\author{Chen-Hui Xie}
	\affiliation{School of Physics, Beijing Institute of Technology, Beijing, 100081, China}
	
	\author{Runyu Lei}
	\affiliation{School of Physics and Astronomy, Beijing Normal University, Beijing 100875, China.}
	
	\author{Jiayi Liu}
	\affiliation{School of Physics, Beijing Institute of Technology, Beijing, 100081, China}
	
	\author{Yihuai Chen}
	\affiliation{School of Physics, Beijing Institute of Technology, Beijing, 100081, China}

	\author{Jinxing Zhang}
	\email{jxzhang@mail.bnu.edu.cn}
	\affiliation{School of Physics and Astronomy, Beijing Normal University, Beijing 100875, China.}

	\author{Yu Gao}
	\email{gaoyu@ihep.ac.cn}
	\affiliation{Institute of High Energy Physics, Chinese Academy of Sciences, Beijing 100049, China}
	
	\author{Sichun Sun}
	\email{sichunssun@bit.edu.cn}
	\affiliation{School of Physics, Beijing Institute of Technology, Beijing, 100081, China}

	\begin{abstract}
		\sloppy{}
		Magnetoelectric phase transitions have been experimentally studied, but no macroscopic theory has been proposed to explain their dynamical origin. In this work, we assume that the axion quasiparticle with frequency $\omega_a$ undergoes a condensation-like process. 
		We show that these magnetoelectric phase transitions can be described within a Ginzburg–Landau (GL) framework by introducing a coupled dynamic parameter, the axion angle, which is proportional to the magnetoelectric coefficient. We derive relations between the static axion angle, the axion frequency, and the phase-transition temperature $T_c$ for different magnetoelectric materials, respectively, and compare these calculations with existing experimental results. We also connect the artificially designed Dzyaloshinskii-Moriya interaction with the axion condensate-like process, so that the relation between the static axion angle and the experimentally measured frequency shift $\Delta f$ can be obtained.
	\end{abstract}
	\maketitle
	
	\section{Introduction}

	Magnetoelectric (ME) effects have attracted considerable attention in condensed matter physics in recent years. Their essential feature is that external electric or magnetic fields induce a coupled response in magnetic and electric degrees of freedom through spin--orbit coupling (SOC) in ME materials \cite{Rivera2009ASR,Fiebig2009CurrentTO,10.21468/SciPostPhys.6.4.046,article,HEHL20081141}.
	The emergence of the ME effect is generally associated with the simultaneous breaking of time-reversal and spatial inversion symmetries \cite{TokuraKawasaki-4,2019arXiv190201532D}.

	For the ME phase transition, most previous studies have focused on the system response near the phase-transition point. Experimental investigations have explored the dependence of the ME coefficient on material volume, material composition, and environmental temperature \cite{LiuSong-3,PhysRevLett.133.156505}. In Ref. \cite{LiuSong-3}, the Dzyaloshinskii-Moriya interaction (DMI) theory is the microscopic generation mechanism for the ME phase transition phenomenon. By introducing the strength of the polarization-induced interfacial DMI, the phase transition temperature $T_c$ can be shown to be proportional to the DMI strength. However, this theory suffers from several limitations. In particular, it cannot describe the frequency dependence of the ME coefficient $\alpha$, nor can it quantitatively describe the relation between the peak value of the ME coefficient and the phase transition temperature $T_c$ \cite{Banerjee:2014hna}.
	Motivated by the interesting connection between the ME coefficient and axion physics, we attempt to employ the extended Ginzburg-Landau (GL) theory to boost the ME coefficient into a dynamic axion field and further describe the relation of the axion quasiparticle (AQ) angle $\theta$, which is proportional to the ME coefficient $\alpha$, to the AQ frequency $\omega_a$. Furthermore, the dynamical axion field goes through a condensation-like transition to trigger the ME phase transition. We then establish a connection between the peak value of the ME coefficient $\alpha$ and the phase transition temperature $T_c$.

	%
	The linear magnetoelectric effect can be expressed as
	$P_i=\alpha_{ij}H_j,
	M_i=\alpha_{ji}E_j,$
	where $P_i$ and $M_i$ denote the polarization and magnetization components, respectively, while $E_j$ and $H_j$ represent the electric and magnetic field components. The ME coupling tensor is defined as 
	$\alpha_{ij}=\left.\frac{\partial P_i}{\partial H_j}\right|_{\mathbf{E}=0}$    \cite{Lei:2025vek}.
	The linear ME coefficient $\alpha=\alpha_{ij}\delta_{ij}$ can be represented by the AQ angle $\theta$, related through
	$\theta=\frac{4\pi^2}{e^2}\alpha$ (natural units), where $\delta_{ij}$ denotes the Keonecker delta and $e$ denotes the elementary charge \cite{cmpk-d882,PhysRevD.110.025014,PhysRevB.109.144304,rf3t-9wfh,LiWang-8}.
	This correspondence was first established in studies of axion electrodynamics in topological magnetic insulators (TMIs) and Weyl semimetals \cite{LiWang-8,sekine_axion_2021,eerenstein_multiferroic_2006, PhysRevLett.102.146805,PhysRevLett.49.405,PhysRevB.78.195424,PhysRevLett.58.1799}.
	In the presence of either time-reversal symmetry $\mathcal{T}$ or inversion symmetry $\mathcal{P}$, the AQ angle $\theta$ is quantized in insulating systems. When both $\mathcal{T}$ and $\mathcal{P}$ symmetries are broken, the AQ angle $\theta$ becomes a continuous variable and therefore exhibits a continuous correspondence with the linear ME coefficient $\alpha$ \cite{qiu_observation_2025}.
	The dynamics of $\theta$ originates from fluctuations of the magnetic order parameter \cite{Lhachemi:2023eha}. In Ref.~\cite{qiu_observation_2025}, coherent oscillations of $\theta$ were identified as a dynamical axion quasiparticle (DAQ). Through Kerr rotation and Berry-curvature measurements in two-dimensional MnBi$_2$Te$_4$, the authors experimentally observed the DAQ signal, providing direct evidence for axion-like collective excitations in ME systems.

	In this work,
	we first state the existing experiments and their main results in \ref{sec:exp}.
	In \ref{sec:BEC}, we suppose AQs in ME materials are a Bose gas, and derive the relationship between the static AQ angle $\theta_0$ and the phase transition temperature $T_c$. We also fit the experimental measurement and draw curves of $\theta_0$ with respect to $T_c$.
	Then, we establish an extended GL description for ME systems and draw the spectrum of $\theta_0$ with respect to AQ frequency $\omega_a$ in \ref{sec:GL2}.
	In \ref{cha:dmi}, we relate the DMI theory and the AQ condensation-like process, giving fitting curves for $\theta_0$ with respect to frequency shift $\Delta f$. Finally, we summarize the main results in \ref{sec:con}. 
	In this paper, we use the natural units convention ($\hbar=c=k_B=1$) unless units are explicitly specified.

	%
	%

	\section{existing experiment }\label{sec:exp}
	
	In reference \cite{LiuSong-3}, an emergent ME phase transition is realized in superlattices with artificially induced ferroelectricity, driven by the interfacial Dzyaloshinskii-Moriya interaction (DMI). Antiferromagnetic Ruddlesden–Popper $\text{Sr}_2\text{IrO}_4$ (SIO) and perovskite paraelectric 
	$\text{Sr}\text{TiO}_3$ (STO) are chosen for epitaxial superlattice fabrication due to their compatible growth conditions \cite{PhysRevLett.133.156505,2008PhRvL.101g6402K,PhysRevLett.101.226402,doi:10.1126/science.1167106,2008JPCM...20C5201K,2013PhRvB..87n0406Y,2019PhRvB..99h5125P,Jackeli:2009qje,PhysRevLett.114.096404,LiuSong-3}. This setup creates a non-equivalent interface that breaks space-inversion symmetry.  Replacing the paraelectric STO with ferroelectric 
	$\text{Ba}\text{TiO}_3$ (BTO) further modulates the interfacial DMI, raising the transition temperature from 46 K to 203 K. Atomic-scale layer-by-layer growth is achievable for 
	$(\text{SIO})_m/(\text{STO})_n (\text{I}_m/\text{T}_n)$ and $(\text{SIO})_m/(\text{BTO})_n (\text{I}_m/\text{B}_n)$ superlattices, where $m$ and $n$ denote the stacking numbers of SIO, STO, and BTO layers, respectively.
	
	%

	\begin{figure}[t]
		\centering
		\includegraphics[width=0.38\textwidth]{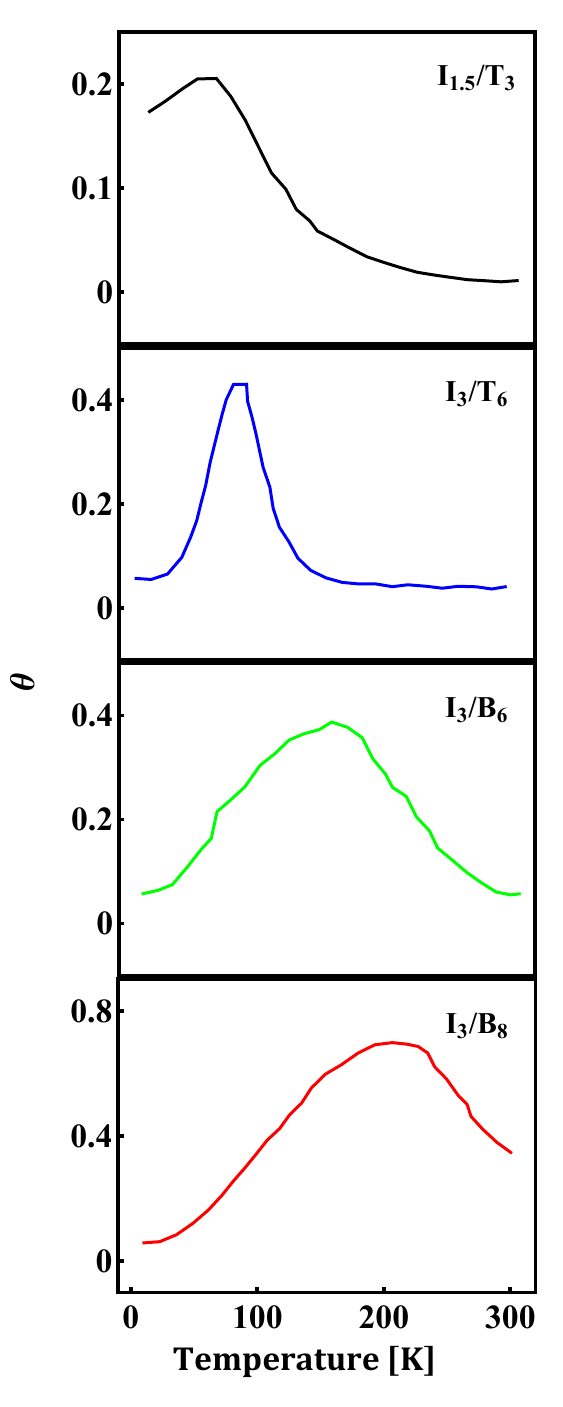}
		\includegraphics[width=0.38\textwidth]{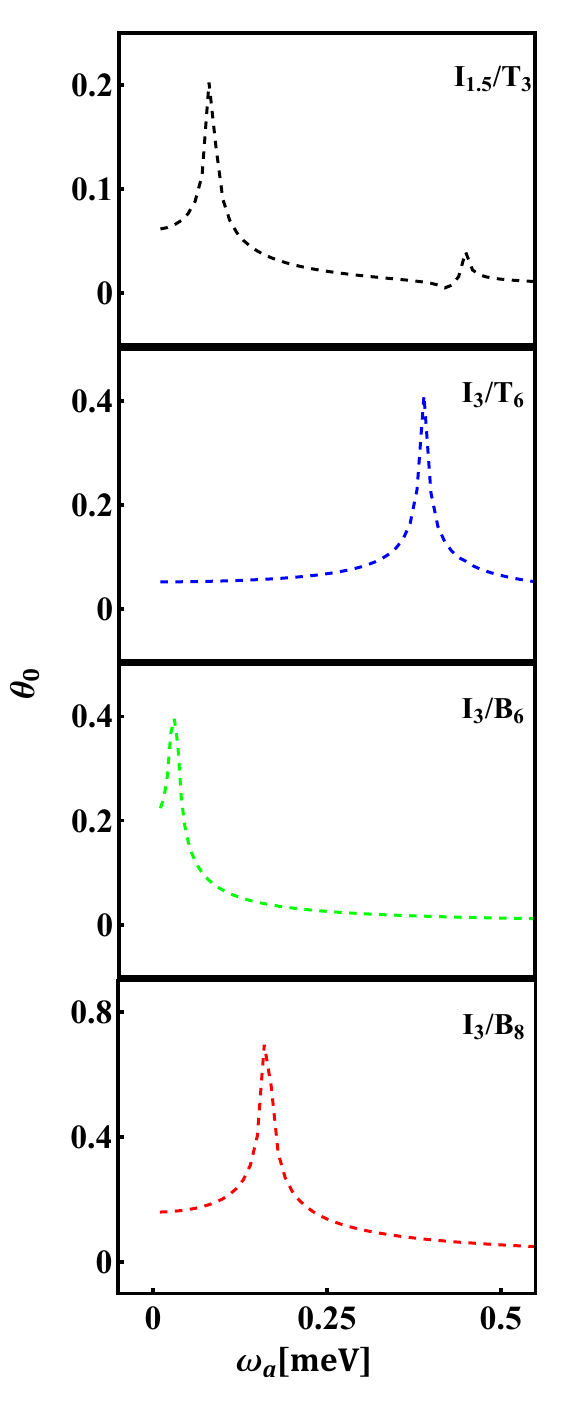}
		\caption{Left: The relationship between the AQ angle $\theta$ and the temperature $T$.  The black, blue, green, and red curves represent cases where the ME phase transition temperature $T_c=$ 46 K, 77 K, 166 K, and 203 K, respectively. These curves are got by $\theta=\frac{4\pi^2}{e^2}\alpha$ with the experimental measurement data of the linear ME coefficient $\alpha$ from Ref. \cite{LiuSong-3}. Right: The spectrum of the static AQ angle $\theta_0$ with respect to the AQ frequency $\omega_a$.  The black, blue, green, and red curves represent cases  $T=T_{C_M}=$ 46 K, 77 K, 166 K, and 203 K, respectively. For each case, the polarization $P_0=0.02, 0.2, 0.55, 0.8~\mu\text{C}/\text{cm}^2$ and magnetization $M_0= 2.2, 2.7, 3.5, 4.2$~emu/cm$^3$, respectively. These curves are got by theoretical calculation and fitting. } 
		\label{fig:thetaT}
	\end{figure}

	In Ref. \cite{LiuSong-3}, authors get the ME phase transition curves for different materials I$_{1.5}$/T$_3$, I$_3$/T$_6$, I$_3$/B$_6$, I$_3$/B$_8$, which can be equally represent by the AQ angle $\theta$ with the relation $\theta=\frac{4\pi^2}{e^2}\alpha$, as shown in the left panel of Fig. \ref{fig:thetaT}. 
	The black, blue, green, and red curves represent cases with magnetic phase transition temperatures $T_{C_M}=$ 46, 77, 166, and 203~K, respectively. For each case, the polarization at the phase transition point is $P_0=0.02, 0.2, 0.55, 0.8~\mu\text{C}/\text{cm}^2$, and the magnetization at the phase transition point is $ M_0=2.2, 2.7, 3.5, 4.2$~emu/cm$^3$, respectively. When the temperature $T=T_{C_M}$, the AQ angle $\theta$ reaches the maximum values, denoted as the static AQ angle $\theta_0= 0.194, 0.423, 0.397, 0.701$, respectively. In the superlattice heterostructures in Ref. \cite{LiuSong-3}, the ME phase transition is triggered by the emergence of magnetic order via interfacial DMI. As a result, the ME phase transition temperature $T_c$ is identical to the magnetic phase transition temperature $T_{C_M}$. We will not distinguish between the two in the subsequent discussion. Next, we fit $\theta_0$ values for different ME materials by constructing a phenomenological theoretical framework. 
	All relevant experimental measurement data are summarized in Table \ref{tab:emd}.

	\section{axion quasiparticle condensation}\label{sec:BEC}
	
	In this section, we suppose that AQ in ME materials may undergo a BEC-like process with a change of temperature \cite{AsteriaZahn-11,GiamarchiRuegg-12,PhysRevB.93.100402}.
	In the zero-velocity limit, the AQ field \(a\) is described by an amplitude $a_0$ and a frequency $\omega_a=m_a$, which leads to the form: $a(t)=a_0e^{-i\omega_at}$. The amplitude is related to AQ density $\rho_a=\frac{1}{2}m_a^2|a_0|^2\label{eq:rho}$, where $m_a$ is the AQ mass \cite{Millar:2016cjp}.
	With $\rho_a=n_a m_a$ and $\theta=ag_{a\gamma}$, where $n_a$ is the AQ number density and $g_{a\gamma}$ is the AQ-photon coupling coefficient, we can get \cite{2009PhRvL.103k1301S,2010AIPC.1274...91Y,2010arXiv1012.1553S,2012PhRvL.108f1304E,2011arXiv1111.3976E}
	\begin{align}
		n_a=\frac{1}{2}m_a|a_0|^2=\frac{1}{2}m_a\left(\frac{\theta_0}{g_{a\gamma}}\right)^2.
	\end{align}
	So we can derive
	\begin{align}
		\theta_0=\sqrt{\frac{2n_a}{m_a}}g_{a\gamma}\label{eq:theta0na}.
	\end{align}

\begin{figure}[t]
		\centering
		\includegraphics[width=0.6\textwidth]{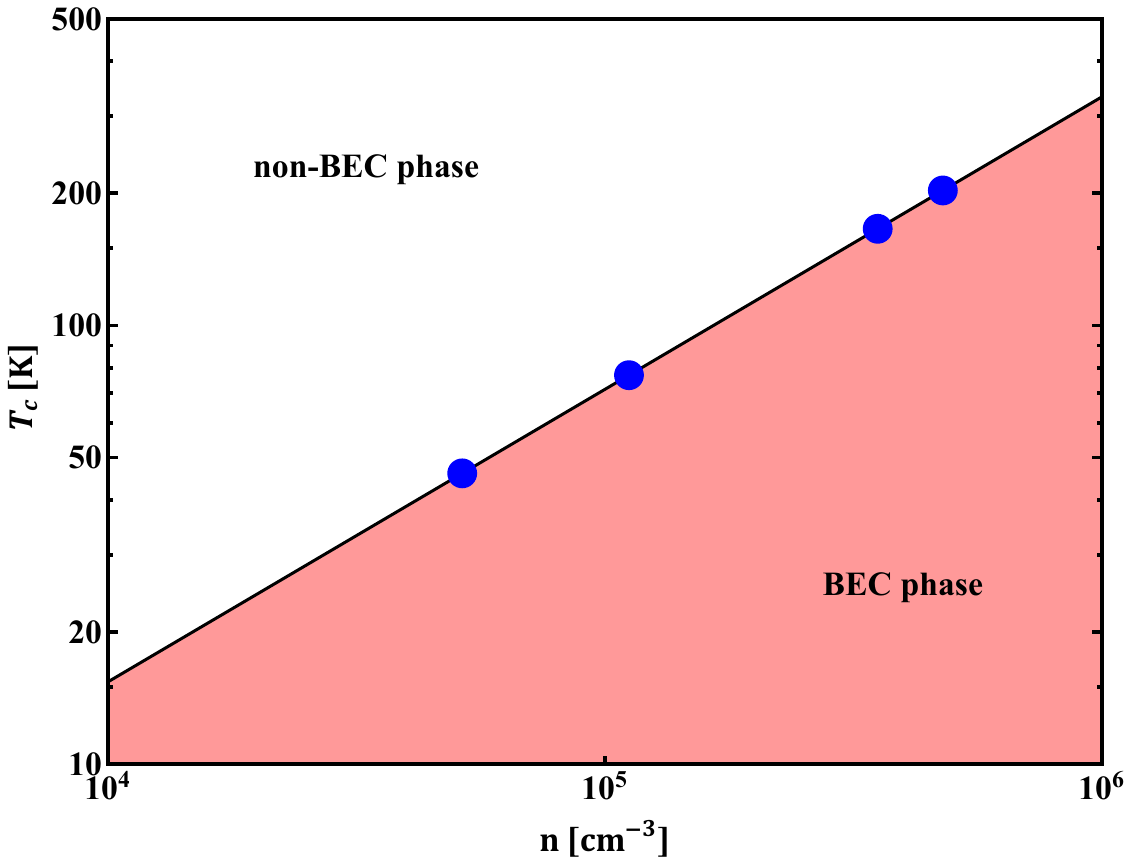}
		\caption{The BEC phase diagram. The blue dots represent cases $T_c$=46, 77, 166, 203~K from the experimental measurement in Ref. \cite{LiuSong-3}.} 
		\label{fig:phasediagram}
	\end{figure}

	Nonrelativistically, the collective excitation described in the subsequent GL framework can be interpreted as a bosonic AQ mode associated with coupled magnetization and polarization fluctuations.
	Near the ME phase transition, the excitation energy becomes low enough that long-wavelength modes dominate the dynamics.
	In this regime, the AQs can be approximately treated as a weakly interacting nonrelativistic Bose gas \cite{PhysRevD.110.025014,SantraBaals-13,Koster_2026,Jalali_Mola_2026,rf3t-9wfh,PhysRevB.109.144304}.
	Therefore, the BEC theory for dilute bosons can be applied phenomenologically to describe the collective condensation behavior of the AQs \cite{SchneiderBracher-14,MoritaYoshioka-15,8wdy-2zbw}.
	The critical temperature for the BEC phase transition in the natural units convention is 
	\begin{align}\label{eq:Tcn2}
		T_c=\frac{2\pi}{\zeta(3/2)^{2/3}}\frac{n_a^{2/3}}{m_a},
	\end{align}
	where  $\zeta(3/2)\approx2.612$ is the Riemann zeta function.
	The BEC phase diagram got from Eq. \ref{eq:Tcn2} is shown in Fig. \ref{fig:phasediagram}.
	With Eq. \ref{eq:theta0na} and Eq. \ref{eq:Tcn2}, we can get

	\begin{align}
		\label{eq:theta0Tc2}\theta_0=\sqrt{2}g_{a\gamma}\left(\frac{\zeta(3/2)^2T_c^3m_a}{\pi^6}\right)^{1/4}.
	\end{align}

	We note that treating AQs as a weakly interacting Bose gas is a macroscopic description rather than a microscopic description. This analogy is adopted to derive a simple power-law scaling between the static AQ angle 
	$\theta_0$ and the ME phase transition temperature $T_c$. The theory fits experimental data near the phase transition point, as we describe the phase transition as a Bose-gas condensation process, although it does not capture magnetization and polarization far from $T_c$.

\begin{figure}[t]
		\centering
		\includegraphics[width=0.6\textwidth]{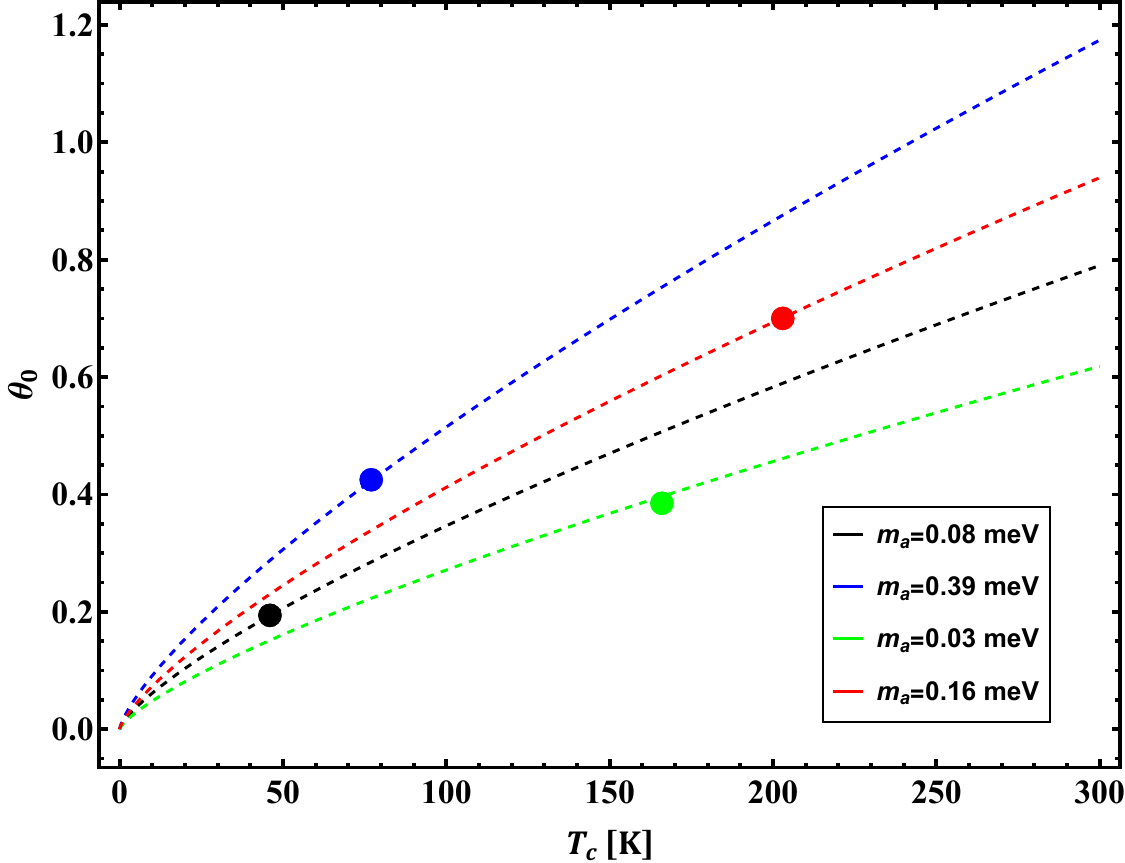}
		\caption{The relationship between the ME coupling $\theta_0$ and $T_c$.  The black, blue, green, and red curves represent cases  $m_a=0.08, 0.39, 0.03, 0.16$~meV for different ME materials I$_{1.5}$/T$_3$, I$_3$/T$_6$, I$_3$/B$_6$ and I$_3$/B$_8$, respectively. The dots are obtained from the experimental measurement in Ref. \cite{LiuSong-3}.} 
		\label{fig:theta0Tc} \end{figure}

	Through Eq. \ref{eq:theta0Tc2}, we can derive the relationship between the static AQ angle $\theta_0$ and phase transition temperature $T_c$, as shown in Fig. \ref{fig:theta0Tc}. Here, we take AQ-photon coupling coefficient $g_{a\gamma}=f^{-1}=0.225~\text{meV}^{-1}$.
	The value of $g_{a\gamma}$
	is not derived from first principles but is chosen as a phenomenological benchmark. This choice is motivated by the requirement that the resulting static effective AQ masses $m_a$
	fall within the meV range, which is physically reasonable for collective excitations in ME systems \cite{qiu_observation_2025}.
	With Eq. \ref{eq:theta0Tc2} and the values of the static AQ angles in different ME materials got from experimental measurement $\theta_0= 0.194, 0.423, 0.397, 0.701$, we can derive the static effective AQ masses $m_a=0.08, 0.39, 0.03, 0.16$~meV for ME materials I$_{1.5}$/T$_3$, I$_3$/T$_6$, I$_3$/B$_6$ and I$_3$/B$_8$, respectively.
	
	It can be observed that for the ME materials and superlattice structures studied in this work, when the basic ME materials are the same, a greater number of layers yields a larger measured $\theta_0$ and a correspondingly larger fitted $m_a$. For instance, Sample I$_3$/T$_6$ exhibits larger $\theta_0$ and $m_a$ values than Sample I$_{1.5}$/T$_3$, while Sample I$_3$/B$_8$ shows larger $\theta_0$ and $m_a$ values than Sample I$_3$/B$_6$. For superlattices with identical layer counts, structures incorporating BTO display smaller $\theta_0$ and $m_a$ than those composed of STO. For example, Sample I$_3$/B$_6$ possesses smaller $\theta_0$ and $m_a$ than Sample I$_3$/T$_6$.

	\section{the extended Ginzburg-Landau Theory}\label{sec:GL2}
	In an isotropic bulk model, we describe the longitudinal dynamics of the internal multiferroic degrees of freedom using an effective approach. We expand the free energy in terms of the order parameters. The free energy density $\mathcal{F}$ is given by $F[\mathbf{M}, \mathbf{P}] = \int \text{d}^3 r~\mathcal{F}[\mathbf{M}, \mathbf{P}]$, and in the absence of external electric and magnetic fields, the formulation is described by \cite{Roising:2021lpv, Lei:2025vek}:
	\begin{align}
		\mathcal{F}[\mathbf{M}, \mathbf{P},\theta] &= \mathcal{F}_M[\mathbf{M}] +\mathcal{F}_P[\mathbf{P}] +\mathcal{F}_\theta[\theta]-\frac{e^2}{4\pi^2}\theta\mathbf{P}\cdot\mathbf{M} , \\
		\label{Eq:FM}\mathcal{F}_M[\mathbf{M}] &= \alpha_M \lvert \partial_t \mathbf{M} \rvert^2 - \beta_M \lvert \mathbf{\nabla} \cdot \mathbf{M} \rvert^2 -  \gamma_M(T-T_{C_M}) \lvert \mathbf{M} \rvert^2 - \lambda_M \lvert \mathbf{M} \rvert^4 + \dots, \\
		\label{Eq:FP}\mathcal{F}_P[\mathbf{P}] &= \alpha_P \lvert \partial_t \mathbf{P} \rvert^2 - \beta_P \lvert \mathbf{\nabla} \cdot \mathbf{P} \rvert^2 - \gamma_P(T-T_{C_P}) \lvert \mathbf{P} \rvert^2 - \lambda_P \lvert \mathbf{P} \rvert^4 + \dots,\\
		\label{Eq:Ftheta}\mathcal{F}_\theta[\theta]&=\frac{1}{2}f^2|\partial_\mu\theta|^2-\frac{1}{2}f^2m_a^2|\theta|^2-\lambda_\theta|\theta|^4+\dots,
	\end{align}
	where $\alpha$'s, $\beta$'s, $\gamma$'s, and $\lambda$'s are phenomenological coefficients. The transition temperatures ${T}_{C_M}$ and ${T}_{C_P}$ denote the magnetic and electric phase transition temperatures, respectively. 
	$f=1/g_{a\gamma}$ is the decay constant of AQ, where $g_{a\gamma}$ is defined as the coupling coefficient of the interaction between the AQs and photons. Here, $m_a$ is the mass of the AQ and $\theta=ag_{a\gamma}$ is the AQ angle, where $a$ represents the AQ field. Conventional GL theory includes only Eqs. \ref{Eq:FM} and \ref{Eq:FP}. To account for the ME phase transition, we introduce the AQ angle $\theta$ as a coupled parameter into the free energy of the GL theory, as shown in Eq. \ref{Eq:Ftheta}.

	The single-particle dynamics of electronic orbitals and spins correspond to ultrafast microscopic processes on the femtosecond timescale, while collective oscillations of magnetically ordered polarizations and AQs constitute slow macroscopic dynamics at the picosecond scale \cite{ArtyukhinDelaney-17, PhysRevMaterials.1.014401}. These two sets of processes have well-separated time scales.
	When constructing the low-energy effective theory, we integrate out all ultrafast electronic degrees of freedom and retain only slow collective fluctuations. The temporal partial derivatives of order parameters $\partial_t \mathbf{M}$, $\partial_t \mathbf{P}$ and $\partial_t \theta$ describe macroscopic collective dynamics, which can be equivalently incorporated into the extended free-energy density functional $\mathcal{F}[\mathbf{M}, \mathbf{P},\theta]$ as effective kinetic terms.
	This framework constitutes a nonequilibrium phenomenological extension of GL theory, distinguishing it from conventional equilibrium GL formalisms {\color{red}{\cite{Wilson2026LightIS,Chen_2024,xue2024case,Harbick_2025}}} .

	The classical equations of motion for M, P, and $\theta$ are derived by 
	\begin{align}
		\frac{\partial \mathcal{F}[\mathbf{M}, \mathbf{P},\theta]}{\partial M}&=\alpha_M\partial^2_t\mathbf{M}-\beta_M\nabla^2\mathbf{M}+\gamma_M(T-T_{C_M})\mathbf{M}+2\lambda_M|\mathbf{M}|^2\mathbf{M}+\frac{e^2}{8\pi^2}\theta\mathbf{P}=0,
		\\
		\frac{\partial \mathcal{F}[\mathbf{M}, \mathbf{P},\theta]}{\partial P}&=\alpha_P\partial^2_t\mathbf{P}-\beta_P\nabla^2\mathbf{P}+\gamma_P(T-T_{C_M})\mathbf{P}+2\lambda_P|\mathbf{P}|^2\mathbf{P}+\frac{e^2}{8\pi^2}\theta\mathbf{M}=0,
		\\
		\frac{\partial \mathcal{F}[\mathbf{M}, \mathbf{P},\theta]}{\partial \theta}&=f^2\partial_\mu\partial^\mu\theta+f^2m_a^2\theta+4\lambda_\theta|\theta|^2\theta+\frac{e^2}{4\pi^2}\mathbf{P}\cdot\mathbf{M}=0.
	\end{align}
	We consider spatially homogeneous solutions here, so that we have $\mathbf{P}(t;x)=P(t)\hat{e}, \mathbf{M}(t;x)=M(t)\hat{e}.$
	We also have $|\partial_\mu\theta|^2=|\partial_t\theta|^2$    and  $\partial_\mu\partial^\mu\theta=\partial_t\partial^t\theta=\ddot{\theta}$, thus we can derive
	\begin{align}
		\label{eq:MthetaP1}	\alpha_M \ddot M +\gamma_M(T-T_{C_M}) M + 2 \lambda_M M^3 +\frac{e^2}{8\pi^2}\theta P\, &= \, 0, \\
		\label{eq:PthetaM1}	\alpha_P \ddot P+\gamma_P(T-T_{C_P}) P + 2 \lambda_P P^3 +\frac{e^2}{8\pi^2}\theta M\, &= \, 0,\\
		\label{eq:thetaPM1}	f^2\ddot{\theta}+f^2m_a^2\theta+4\lambda_\theta\theta^3+\frac{e^2}{4\pi^2}PM&=0.
	\end{align}

	The $\lambda_MM^3$ term in Eq. \ref{eq:MthetaP1} is also called the cubic Gross-Pitaevskii (GP) term. The cubic GP term, which breaks time-reversal symmetry $\mathcal{T}$, is often used to describe BEC behavior, related to GP equations \cite{Jalali_Mola_2026}. Because of this deep connection between BEC and the ME phase transition, we naturally think that the BEC behavior of AQ may give a phenomenologically effective description for the ME phase transition. Thus, it is consistent with our treatment of AQs as bosons and the ME phase transition as a BEC process in the former section.

	We set $M=M_0+\delta M, P=P_0+\delta P$ and $\theta=\theta_0+\delta\theta$, where $M_0$ and $P_0$ are the values of the static polarization and magnetization when the temperature equals the phase transition temperature, and $\theta_0$ is the static AQ angle. We employ the oscillator solutions $\delta\theta(t)=\delta\theta~e^{-i\omega_at}$.
	Through the classical equations of motion for M, P, and $\theta$, we can get the relation between the parameters in the above equations
	\begin{align}\label{Eq:theta0wa1}
		{12\lambda_\theta}\theta_0^2+\frac{e^4}{2\pi^4}\frac{4\pi^2(P_0^2F_P(\omega_a)+M_0^2F_M(\omega_a))-e^2\theta_0P_0M_0}{e^4\theta_0^2-64\pi^2F_M(\omega_a)F_P(\omega_a)}-i\omega_af^2\eta_\theta=0,
	\end{align}
	where we define functions $F_M(\omega_a)\equiv i \alpha_M\eta_M \omega_a-\alpha_M\omega_a^2+6 \lambda_M M_0^2$ and $F_P(\omega_a)\equiv i\alpha_P \eta_P \omega_a-\alpha_P\omega_a^2+m_P^2$.
	Through Eqs. \ref{eq:theta0Tc2} and \ref{Eq:theta0wa1}, we can derive the relationship between static AQ angle $\theta_0$ and AQ frequency $\omega_a$ when $T=T_c$, shown as the right panel of Fig. \ref{fig:thetaT}. Here, we take phenomenological coefficient $\lambda_M=$ 1.5, 22, 0.08, 1.6~kOe$^{-2}$, respectively. We take the phenomenological coefficient $\alpha_M=1$~meV$^{-2}$.
	and $\alpha_P = 10^{-2}~\text{meV}^{-2}$ . $\eta_M=5~\mu\text{eV}$ and $\eta_P=5~\mu\text{eV}$ are damping coefficients \cite{Roising:2021lpv}. $m_P=4.5\times10^{-2}$  is the effective mass at equilibrium.
	$\gamma_M=10^{-5}~\text{K}^{-1}$ is a phenomenological coefficient.
	Here, we set $\lambda_\theta= 0.005, 0.0315, 2, 0.3$~kOe$^2$ for cases  $T_{c}=$ 46, 77, 166, 203~K.
	From the right panel of Fig. \ref{fig:thetaT}, we can see that the static AQ angle $\theta_0$ reaches its peak when the critical AQ frequency $\omega_a^0=m_a\approx0.08, 0.39, 0.03, 0.16$~meV, respectively. The corresponding peak values of $\theta_0$ are $(\theta_0)_\text{max}\approx$ 0.196, 0.410, 0.395, 0.694, respectively.
	All phenomenological coefficients are constrained by the measured static polarization 
	$P_0$, magnetization $M_0$, phase transition temperature $T_c$
	and experimentally measured ME coupling coefficient $\theta$ from reference experiments. 
	In summary, we adapt the values of $\lambda_M$ and $\lambda_\theta$ to draw the spectrum of the static AQ angle $\theta_0$ with respect to AQ frequency $\omega_a$, realizing that the $\theta_0$ reaches its peak when $\omega_a=\omega_a^0=m_a$ while the $(\theta_0)_{\text{max}}$ is near the $\theta_0$ from experimental measurement for each case.
	Tables \ref{tab:fp} and \ref{tab:cp} summarize all relevant fitting and cited parameters, respectively.
	

	The magnitude of the phenomenological coefficient $\lambda_M$ in front of the quartic term of magnetization $M$
	in the free energy density $\mathcal{F}$ significantly affects the value of $\omega_{a}^0$. Within a certain range, the larger $\lambda_M$ is, the greater $\omega_{a}^0$ becomes, while $\theta_0$ may decrease. At the same time, the phenomenological coefficient $\lambda_\theta$ in front of the quartic term of AQ $\theta$ also exerts an influence on $\theta_0$. A larger $\lambda_\theta$ leads to a smaller $\theta_0$. We leave the further explanation of these parameters for future work.

	The peak values of static AQ angles $(\theta_0)_{\text{max}}$ solved from Eq. \ref{Eq:theta0wa1} and the values of $\theta_0$ from the experimental measurement are almost equal under the set of static effective AQ masses $\omega_a^0=m_a$ got from Eq. \ref{eq:theta0Tc2}. This consistency verifies the internal self-consistency of our theoretical framework, and the proposed theoretical framework gives a frequency perspective on the ME phase transition.

	\section{Dzyaloshinskii-Moriya interaction}
	\label{cha:dmi}
	To achieve a better integration between theory and experiment, we incorporate the AQ condensation theory in Section \ref{sec:BEC}, the DMI theory, and the Brillouin light scattering (BLS) measurements within a unified theoretical framework.
	In the ME materials in Ref. \cite{LiuSong-3}, the DMI serves as the microscopic origin of the ME effect \cite{PhysRevLett.97.167204,PhysRevB.94.064418,PhysRevB.85.224413,OkamuraKagawa-10,PhysRevLett.113.107203,s9l2-m9tt,b6gt-w5xy}.
	The ME phase transition emerges from the interplay between the spin-orbit coupling (SOC) and the broken inversion symmetry at the interface \cite{ManchonKoo-16}. We begin by constructing the effective spin Hamiltonian that governs the magnetic interactions \cite{LiuSong-3,2014PhRvX...4c1045B}:
\begin{align}
	\mathcal{H} &= \underbrace{-J_{\text{eff}} \sum_{i,\mu} \boldsymbol{S}_i \cdot \boldsymbol{S}_{i+\hat{\mu}}}_{\text{Heisenberg coupling}}
	-\underbrace{A_c \sum_{i} \left( \boldsymbol{S}_i^y \boldsymbol{S}_{i+\hat{x}}^y + \boldsymbol{S}_i^x \boldsymbol{S}_{i+\hat{y}}^x \right) }_{\text{Compass anisotropy}} 
	-\underbrace{D \sum_{i} \left[ \hat{\boldsymbol{y}} \cdot \left( \boldsymbol{S}_i \times \boldsymbol{S}_{i+\hat{x}} \right) - \hat{\boldsymbol{x}} \cdot \left( \boldsymbol{S}_i \times \boldsymbol{S}_{i+\hat{y}} \right) \right]}_{\text{Interfacial DMI}},
\end{align}
where $\hat{x}$ and $\hat{y}$ are unitvectors and $\hat{\mu}=\hat{x},\hat{y}$. $\mathbf{S}_i$ and $\mathbf{S}_{i+\hat{\mu}}$ are the magnetic moment at the site $i$ and ${i+\hat{\mu}}$. The coupling parameters \( J_{\text{eff}} \), \( A_c \), and \( D \) are fundamentally related through the SOC strength \( \lambda \) and hopping parameter \( t \). The trigonometric relations \( J_{\text{eff}} = \tilde{J} \cos 2\delta \), \( A_c = \tilde{J}(1 - \cos 2\delta) \), and \( D = \tilde{J} \sin 2\delta \) originate from the angular parameter \( \delta \) defined by \( \tan \delta = \lambda/t \), with \( \tilde{J} \propto \sqrt{t^2 + \lambda^2} \). With \( \lambda \ll t \), the small-angle approximation yields \( D \approx 2\delta \tilde{J} \approx 2\lambda \), revealing the direct proportionality between DMI strength and SOC \cite{2025arXiv250417772L,2014PhRvX...4c1045B,2025arXiv250602192G,2015NatMa..14..871M,2024arXiv241201631Y}.

\begin{figure}[t]
	\centering
	\includegraphics[width=0.6\textwidth]{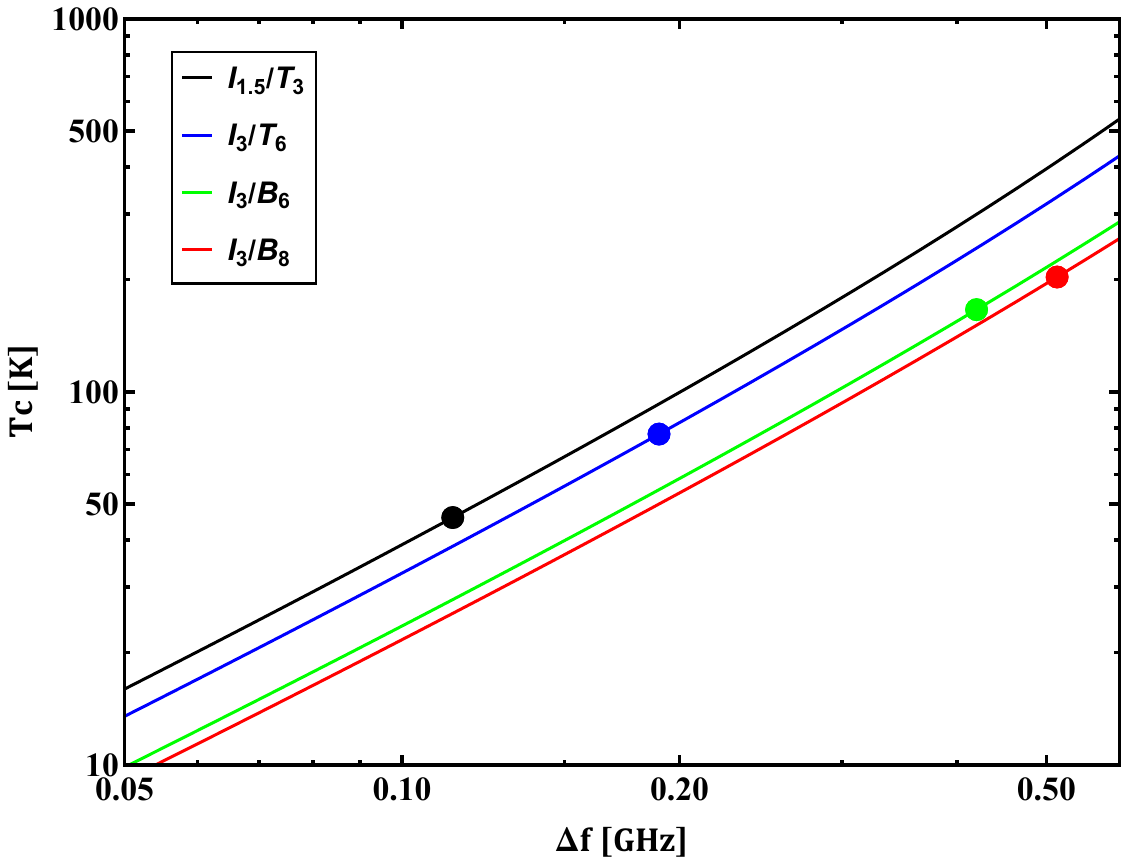}
	\caption{The fitting curves for the phase transition temperature $T_c$ and frequency shift \( \Delta f \).  The black, blue, green, and red curves represent cases  I$_{1.5}$/T$_3$, I$_3$/T$_6$, I$_3$/B$_6$, and I$_3$/B$_8$, respectively. The blue and green dots are experimental data, while the black and red dots are estimates.} 
	\label{fig:Tcdeltaf}
\end{figure}

\begin{figure}[t]
	\centering
	\includegraphics[width=0.6\textwidth]{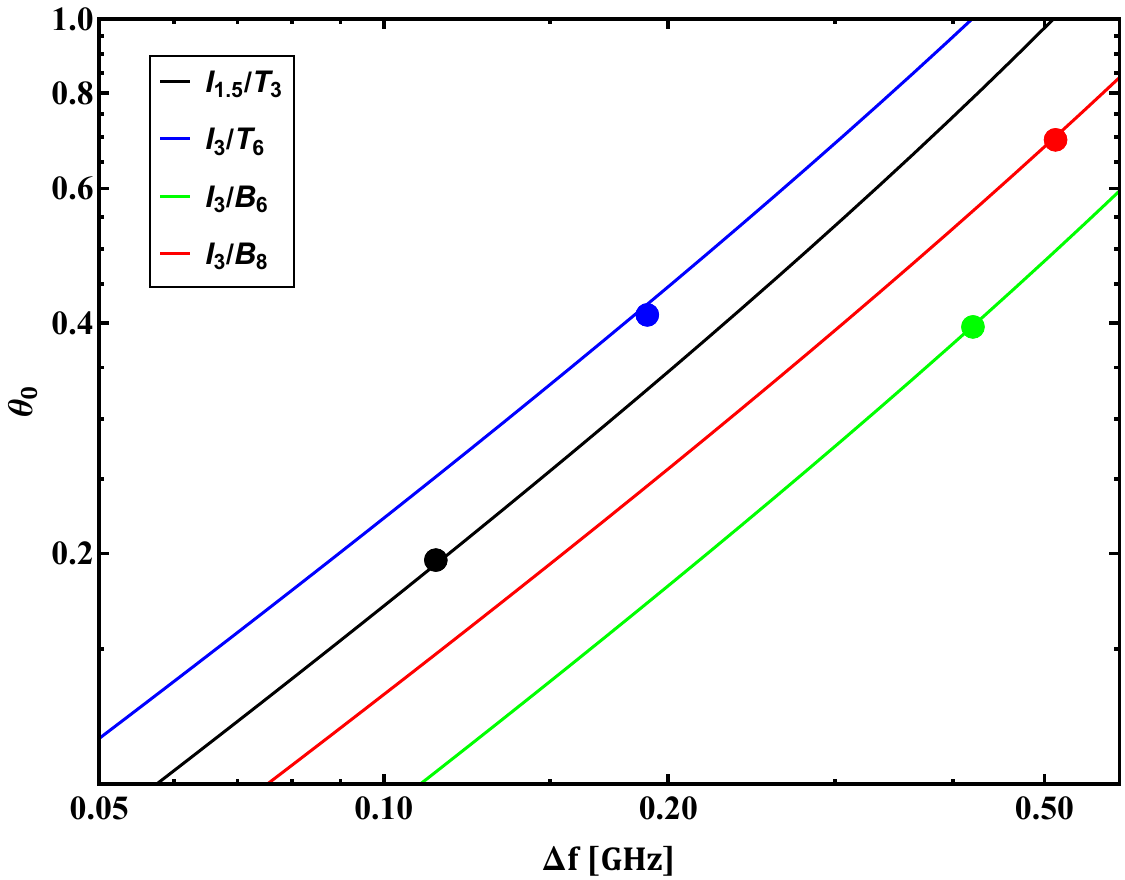}
	\caption{The fitting curves for the static AQ angle $\theta_0$ and the frequency shift \( \Delta f \).  The black, blue, green, and red curves represent cases  I$_{1.5}$/T$_3$, I$_3$/T$_6$, I$_3$/B$_6$, and I$_3$/B$_8$, respectively. The blue and green dots come from experimental data, while the black and red dots come from estimation.} 
	\label{fig:theta0deltaf}
\end{figure}

The critical temperature \( T_c \) for the phase transition can be derived by examining the free energy landscape of the quasi-two-dimensional magnetic system. From the Mermin-Wagner theory adapted for symmetry-broken systems, the free energy difference between ordered and disordered phases is given by \cite{2014PhRvX...4c1045B}
\begin{align}
	\Delta F = 4\pi J_{\text{eff}} - k_B T_c \ln\left( \frac{\beta}{\lambda} \right),
\end{align}
where \( \beta \) is a phenomenological parameter accounting for higher-order interactions. At the phase transition boundary \( \Delta F = 0 \), we can get
\begin{align}\label{Eq:dF=0}
	k_B T_c =4\pi J_{\text{eff}} \left[ \ln\left( \frac{\beta}{\lambda} \right) \right]^{-1}.
\end{align}
Substituting \( J_{\text{eff}} = \tilde{J} \cos 2\delta \) into Eq. \ref{Eq:dF=0} and recognizing that \( 2\lambda \simeq D \) through the earlier approximation, we can obtain the function of the phase transition temperature $T_c$ with respect to the DMI strength $D$ \cite{10.1143/PTP.63.387}
\begin{align}
	T_c =4\pi\frac{D}{2\delta k_B}\left[\text{ln}\left(\frac{2\beta}{D}\right)\right]^{-1}.\label{eq:TcD}
\end{align}
Here, we use $\text{cos}\ 2\delta\approx1$ with the small-angle approximation. This logarithmic dependence demonstrates that the phase transition temperature $T_c$ scales monotonically with the DMI strength $D$, albeit with diminishing returns at higher \( D \) values due to the logarithmic denominator. The theoretical prediction is quantitatively confirmed by Brillouin light scattering (BLS) measurements, where the observed frequency shift \( \Delta f \) between Stokes and anti-Stokes peaks directly quantifies the strength of the interfacial DMI \cite{LiuSong-3,DuZhang-9}
\begin{align}
	D_{\text{int}} = \frac{\pi \Delta f M_s}{2\gamma k},\label{eq:Ddeltaf}
\end{align}
where  $\gamma$ represents the gyromagnetic ratio, $M_s$ is the saturation magnetization and $k$ is the module of the wave vector $\mathbf{k}$. 
Here,  the DMI strength $D$ and the strength of the interfacial DMI $D_{\text{int}}$ have the relation
$D_{\text{int}}=D/S$, where $S=a^2$ is the unit cell area and $a\approx0.35~$nm is the lattice constant.
The DMI manifests as a frequency splitting \(\Delta f = f(\mathbf{k}) - f(-\mathbf{k})\) in BLS spectra.
Substituting Eq. \ref{eq:Ddeltaf} into Eq. \ref{eq:TcD}, we can obtain the function of phase transition temperature $T_c$ with respect to frequency shift $\Delta f$
\begin{align}\label{Eq:Tcdeltaf}
	T_c=\frac{\pi^2\Delta fM_sS}{\gamma\delta k_Bk}\left[\text{ln}\left(\frac{4\beta\gamma k}{\pi\Delta fM_sS}\right)\right]^{-1},
\end{align}
where phenomenological parameter $\beta=1$~meV and $\delta=0.05$.
Ref. \cite{LiuSong-3} gives the frequency splitting $\Delta f =0.19~$GHz for ME material  I$_3$/T$_6$ and $\Delta f =0.42~$GHz for ME material  I$_3$/B$_6$. From Eqs. \ref{eq:TcD}  and \ref{Eq:Tcdeltaf}, we can see that $T_c$ is proportional to $D$ or $\Delta f$ when $D$ or $\Delta f$ is sufficiently small, which is supported by the experimental data: $0.42~$GHz/0.19~GHz$\approx$166K/77K.
Thus, we can use $T_c$ to estimate the frequency splitting $\Delta f \approx 0.114$~GHz for I$_{1.5}$/T$_3$ and $\Delta f \approx 0.514$~GHz for I$_3$/B$_8$ by linear scaling of $T_c$ with $\Delta f$ based on the two given experimental points.
With the values of $\Delta f$, the gyromagnetic ratio $\gamma=1.76\times10^{11}$ rad$\cdot$s$^{-1}\cdot$T$^{-1}$ and the saturation magnetization $M_s=2\times10^6$~A/m, we can get $k$ for different ME materials by Eq. \ref{Eq:Tcdeltaf}. The derived  $k\approx16.2, 18.5, 23.8 ,25.4$~rad$\cdot\mu$m$^{-1}$ for materials I$_{1.5}$/T$_3$, I$_3$/T$_6$, I$_3$/B$_6$, I$_3$/B$_8$, respectively. With these values of $k$, we can get the fitting curves for the phase transition temperature $T_c$ and frequency shift $\Delta f$, as shown in Fig. \ref{fig:Tcdeltaf}.
The corresponding strength of the interfacial DMI $D_{\text{int}}$ are $0.13, 0.18, 0.32, 0.36$~mJ/m$^2$. 
By Eqs. \ref{eq:theta0Tc2} and \ref{Eq:Tcdeltaf}, with the values of $m_a$ from Section \ref{sec:BEC}, we can get the fitting curves for axion angle $\theta_0$ and frequency shift \( \Delta f \), shown as Fig. \ref{fig:theta0deltaf}
\cite{PhysRevX.4.031045,10.1143/PTP.63.387}. 
Tables \ref{tab:fp} and \ref{tab:cp} summarize all relevant fitting and cited parameters, respectively.



\section{conclusion}\label{sec:con}


Firstly, by supposing the AQs as a weak-interaction Bose gas, we relate the ME phase transition to their condensation behavior and obtain a relation between the static AQ angle $\theta_0$ and the phase transition temperature $T_c$. By fitting the experimental measurement data, we get the static effective AQ masses $m_a$ for different ME materials.
		Second, this work shows that the ME phase transition can be described phenomenologically by combining GL theory with a new parameter $\theta$ and an AQ condensate-like process. The extended GL model qualitatively gives the dependence of the static AQ angle $\theta_0$ on the AQ frequency $\omega_a$. With the static effective AQ masses $m_a$, we draw the spectrum of the static AQ angle $\theta_0$ with respect to the AQ frequency $\omega_a$.
		Third, we connect the microscopic generation mechanism of the ME phase transition, the DMI theory, with the assumption of an AQ condensate-like process and plot the static AQ angle $\theta_0$ versus the frequency shift $\Delta f$. 
		In summary, we get the dependence of the static AQ angle $\theta_0$ on the phase transition temperature $T_c$,  the AQ frequency $\omega_a$, and the frequency shift $\Delta f$.

The theoretical framework developed in this work still needs further refinement. Firstly, the AQ condensation model for interpreting ME phase transitions requires further experimental verification to confirm its reliability. Specifically, future experiments can refer to Ref. \cite{qiu_observation_2025} to measure the time-domain oscillations of the AQ, then perform a fast Fourier transform (FFT) to transform into the frequency domain, extract the mass of the AQ, and substitute it into Eq. \ref{eq:theta0Tc2} for verification.
Secondly, this work provides a feasible phenomenological framework to describe and fit existing ME experimental data. Further microscopic parameter extraction via first-principles calculations across different material compositions is necessary to fully test the theory's reliability and enable it to predict unmeasured experimental curves. Such microscopic calibration is beyond the scope of the current phenomenological study and will be implemented in follow-up independent research.
Third, the ME coefficient, magnetization, and polarization intensity also depend on the material's thickness and volume, which the present model does not account for, indicating that this phenomenological interpretation still requires further improvement. We intend to discuss this in detail in future work.

\medskip
\textbf{Acknowledgements} \par 

The authors thank Fei Gao for helpful communications, and Xin Liu for providing the dataset for analysis in this work. This work is supported by the National Natural Science Foundation of China (No.  52225205, J.Z, No. T2350005,  No. 12105013 and No. 12447105), the National Key Research and Development Program of China (No. 2023YFA1406500, J.Z and No. 2021YFA0718700, J.Z), the Fundamental Research Funds for the Central Universities (J.Z) and the Beijing Natural Science Foundation (Z240008, J.Z).
\medskip

\bibliographystyle {unsrt}
\bibliography{ref.bib}

\begin{thebibliography}{10}

\bibitem{Rivera2009ASR}
J.~P. Rivera.
\newblock A short review of the magnetoelectric effect and related experimental
  techniques on single phase (multi-) ferroics.
\newblock {\em The European Physical Journal B}, 71:299--313, 2009.

\bibitem{Fiebig2009CurrentTO}
Manfred Fiebig and Nicola~A. Spaldin.
\newblock Current trends of the magnetoelectric effect.
\newblock {\em The European Physical Journal B}, 71:293--297, 2009.

\bibitem{10.21468/SciPostPhys.6.4.046}
N.~P. Armitage and Liang Wu.
\newblock {On the matter of topological insulators as magnetoelectrics}.
\newblock {\em SciPost Phys.}, 6:046, 2019.

\bibitem{article}
Sinisa Coh, David Vanderbilt, Andrei Malashevich, and Ivo Souza.
\newblock Chern-simons orbital magnetoelectric coupling in generic insulators.
\newblock {\em Phys. Rev. B}, 83:085108, Feb 2011.

\bibitem{HEHL20081141}
Friedrich~W. Hehl, Yuri~N. Obukhov, Jean-Pierre Rivera, and Hans Schmid.
\newblock Relativistic analysis of magnetoelectric crystals: Extracting a new
  4-dimensional p odd and t odd pseudoscalar from cr2o3 data.
\newblock {\em Physics Letters A}, 372(8):1141--1146, 2008.

\bibitem{TokuraKawasaki-4}
Yoshinori Tokura, Masashi Kawasaki, and Naoto Nagaosa.
\newblock Emergent functions of quantum materials.
\newblock {\em Nature Physics}, 13(11):1056--1068, 2017.

\bibitem{2019arXiv190201532D}
Shuai {Dong}, Hongjun {Xiang}, and Elbio {Dagotto}.
\newblock {Magnetoelectricity in Multiferroics: a Theoretical Perspective}.
\newblock {\em arXiv preprint arXiv:1902.01532}, February 2019.

\bibitem{LiuSong-3}
Xin Liu, Wenjie Song, Mei Wu, Yuben Yang, Ying Yang, Peipei Lu, Yinhua Tian,
  Yuanwei Sun, Jingdi Lu, Jing Wang, Dayu Yan, Youguo Shi, Nian~Xiang Sun,
  Young Sun, Peng Gao, Ka~Shen, Guozhi Chai, Supeng Kou, Ce-Wen Nan, and
  Jinxing Zhang.
\newblock Magnetoelectric phase transition driven by interfacial-engineered
  dzyaloshinskii-moriya interaction.
\newblock {\em Nature Communications}, 12(1):5453, 2021.

\bibitem{PhysRevLett.133.156505}
Xin Liu, Ting Hu, Yujun Zhang, Xueli Xu, Runyu Lei, Biao Wu, Zongwei Ma, Peng
  Lv, Yuelin Zhang, Shih-Wen Huang, Jialu Wu, Jing Ma, Jiawang Hong, Zhigao
  Sheng, Chenglong Jia, Erjun Kan, Ce-Wen Nan, and Jinxing Zhang.
\newblock {Flexomagnetoelectric Effect in ${\mathrm{Sr}}_{2}{\mathrm{IrO}}_{4}$
  Thin Films}.
\newblock {\em Phys. Rev. Lett.}, 133:156505, Oct 2024.

\bibitem{Banerjee:2014hna}
Sumilan Banerjee, James Rowland, Onur Erten, and Mohit Randeria.
\newblock {Skyrmions in two-dimensional chiral magnets}.
\newblock {\em Phys. Rev. X}, 4(3):031045, 2014.

\bibitem{Lei:2025vek}
Runyu Lei, Chen-Hui Xie, Jiayi Liu, Zhong Liu, Xin Liu, Yu~Gao, Sichun Sun, and
  Jinxing Zhang.
\newblock {Detecting Axion Dark Matter by Artificial Magnetoelectric
  Materials}.
\newblock {\em Annalen Phys.}, 537(12):e00202, 2025.

\bibitem{cmpk-d882}
Yiliang Fan, Rongxiang Zhu, Tongshuai Zhu, Jianzhou Zhao, Huaiqiang Wang, and
  Haijun Zhang.
\newblock Switchable axionic magnetoelectric effect via spin-flop transition in
  topological antiferromagnets.
\newblock {\em Phys. Rev. B}, 113:195137, May 2026.

\bibitem{PhysRevD.110.025014}
Claudio Corian\`o, Mario Cret\`{\i}, Stefano Lionetti, and Riccardo Tommasi.
\newblock Axionlike quasiparticles and topological states of matter: Finite
  density corrections of the chiral anomaly vertex.
\newblock {\em Phys. Rev. D}, 110:025014, Jul 2024.

\bibitem{PhysRevB.109.144304}
M.~Nabil~Y. Lhachemi and Ion Garate.
\newblock Phononic dynamical axion in magnetic dirac insulators.
\newblock {\em Phys. Rev. B}, 109:144304, Apr 2024.

\bibitem{rf3t-9wfh}
Daniel Boyanovsky.
\newblock Probing dynamical axion quasiparticles with two-photon correlations.
\newblock {\em Phys. Rev. B}, 112:174301, Nov 2025.

\bibitem{LiWang-8}
Rundong Li, Jing Wang, Xiao-Liang Qi, and Shou-Cheng Zhang.
\newblock Dynamical axion field in topological magnetic insulators.
\newblock {\em Nature Physics}, 6(4):284--288, 2010.

\bibitem{sekine_axion_2021}
Akihiko Sekine and Kentaro Nomura.
\newblock Axion {Electrodynamics} in {Topological} {Materials}.
\newblock {\em J. Appl. Phys.}, 129(14):141101, 2021.

\bibitem{eerenstein_multiferroic_2006}
W.~Eerenstein, N.~D. Mathur, and J.~F. Scott.
\newblock Multiferroic and magnetoelectric materials.
\newblock {\em Nature}, 442(7104):759--765, August 2006.

\bibitem{PhysRevLett.102.146805}
Andrew~M. Essin, Joel~E. Moore, and David Vanderbilt.
\newblock Magnetoelectric polarizability and axion electrodynamics in
  crystalline insulators.
\newblock {\em Phys. Rev. Lett.}, 102:146805, Apr 2009.

\bibitem{PhysRevLett.49.405}
D.~J. Thouless, M.~Kohmoto, M.~P. Nightingale, and M.~den Nijs.
\newblock Quantized hall conductance in a two-dimensional periodic potential.
\newblock {\em Phys. Rev. Lett.}, 49:405--408, Aug 1982.

\bibitem{PhysRevB.78.195424}
Xiao-Liang Qi, Taylor~L. Hughes, and Shou-Cheng Zhang.
\newblock Topological field theory of time-reversal invariant insulators.
\newblock {\em Phys. Rev. B}, 78:195424, Nov 2008.

\bibitem{PhysRevLett.58.1799}
Frank Wilczek.
\newblock Two applications of axion electrodynamics.
\newblock {\em Phys. Rev. Lett.}, 58:1799--1802, May 1987.

\bibitem{qiu_observation_2025}
Jian-Xiang Qiu, Barun Ghosh, Jan Schütte-Engel, Tiema Qian, Michael Smith,
  Yueh-Ting Yao, Junyeong Ahn, Yu-Fei Liu, Anyuan Gao, Christian Tzschaschel,
  Houchen Li, Ioannis Petrides, Damien Bérubé, Thao Dinh, Tianye Huang,
  Olivia Liebman, Emily~M. Been, Joanna~M. Blawat, Kenji Watanabe, Takashi
  Taniguchi, Kin~Chung Fong, Hsin Lin, Peter~P. Orth, Prineha Narang, Claudia
  Felser, Tay-Rong Chang, Ross McDonald, Robert~J. McQueeney, Arun Bansil, Ivar
  Martin, Ni~Ni, Qiong Ma, David J.~E. Marsh, Ashvin Vishwanath, and Su-Yang
  Xu.
\newblock Observation of the axion quasiparticle in {2D} {MnBi2Te4}.
\newblock {\em Nature}, April 2025.

\bibitem{Lhachemi:2023eha}
M.~Nabil~Y. Lhachemi and Ion Garate.
\newblock {Phononic dynamical axion in magnetic Dirac insulators}.
\newblock {\em Phys. Rev. B}, 109(14):144304, 2024.

\bibitem{2008PhRvL.101g6402K}
B.~J. {Kim}, Hosub {Jin}, S.~J. {Moon}, J.~Y. {Kim}, B.~G. {Park}, C.~S.
  {Leem}, Jaejun {Yu}, T.~W. {Noh}, C.~{Kim}, S.~J. {Oh}, J.~H. {Park},
  V.~{Durairaj}, G.~{Cao}, and E.~{Rotenberg}.
\newblock {Novel J$_{eff}$=1/2 Mott State Induced by Relativistic Spin-Orbit
  Coupling in Sr$_{2}$IrO$_{4}$}.
\newblock {\em Phys. Rev. Lett.}, 101(7):076402, August 2008.

\bibitem{PhysRevLett.101.226402}
S.~J. Moon, H.~Jin, K.~W. Kim, W.~S. Choi, Y.~S. Lee, J.~Yu, G.~Cao, A.~Sumi,
  H.~Funakubo, C.~Bernhard, and T.~W. Noh.
\newblock {Dimensionality-Controlled Insulator-Metal Transition and Correlated
  Metallic State in $5d$ Transition Metal Oxides
  ${\mathrm{Sr}}_{n+1}{\mathrm{Ir}}_{n}{\mathrm{O}}_{3n+1}$ ($n=1$, 2, and
  $\ensuremath{\infty}$)}.
\newblock {\em Phys. Rev. Lett.}, 101:226402, Nov 2008.

\bibitem{doi:10.1126/science.1167106}
B.~J. Kim, H.~Ohsumi, T.~Komesu, S.~Sakai, T.~Morita, H.~Takagi, and T.~Arima.
\newblock Phase-sensitive observation of a spin-orbital mott state in
  {${\mathrm{Sr}}_{2}{\mathrm{Ir}}{\mathrm{O}}_{4}$}.
\newblock {\em Science}, 323(5919):1329--1332, 2009.

\bibitem{2008JPCM...20C5201K}
Y.~{Klein} and I.~{Terasaki}.
\newblock {Insight on the electronic state of Sr$_{2}$IrO$_{4}$ revealed by
  cationic substitutions}.
\newblock {\em Journal of Physics Condensed Matter}, 20(29):295201, July 2008.

\bibitem{2013PhRvB..87n0406Y}
Feng {Ye}, Songxue {Chi}, Bryan~C. {Chakoumakos}, Jaime~A. {Fernandez-Baca},
  Tongfei {Qi}, and G.~{Cao}.
\newblock {Magnetic and crystal structures of Sr$_{2}$IrO$_{4}$: A neutron
  diffraction study}.
\newblock {\em Phys. Rev. B}, 87(14):140406, April 2013.

\bibitem{2019PhRvB..99h5125P}
J.~{Porras}, J.~{Bertinshaw}, H.~{Liu}, G.~{Khaliullin}, N.~H. {Sung}, J.~W.
  {Kim}, S.~{Francoual}, P.~{Steffens}, G.~{Deng}, M.~Moretti {Sala},
  A.~{Efimenko}, A.~{Said}, D.~{Casa}, X.~{Huang}, T.~{Gog}, J.~{Kim},
  B.~{Keimer}, and B.~J. {Kim}.
\newblock {Pseudospin-lattice coupling in the spin-orbit Mott insulator
  Sr$_{2}$IrO$_{4}$}.
\newblock {\em Phys. Rev. B}, 99(8):085125, February 2019.

\bibitem{Jackeli:2009qje}
G.~Jackeli and G.~Khaliullin.
\newblock {Mott Insulators in the Strong Spin-Orbit Coupling Limit: From
  Heisenberg to a Quantum Compass and Kitaev Models}.
\newblock {\em Phys. Rev. Lett.}, 102(1):017205, 2009.

\bibitem{PhysRevLett.114.096404}
D.~H. Torchinsky, H.~Chu, L.~Zhao, N.~B. Perkins, Y.~Sizyuk, T.~Qi, G.~Cao, and
  D.~Hsieh.
\newblock Structural distortion-induced magnetoelastic locking in
  {${\mathrm{Sr}}_{2}{\mathrm{IrO}}_{4}$} revealed through nonlinear optical
  harmonic generation.
\newblock {\em Phys. Rev. Lett.}, 114:096404, Mar 2015.

\bibitem{AsteriaZahn-11}
Luca Asteria, Henrik~P. Zahn, Marcel~N. Kosch, Klaus Sengstock, and Christof
  Weitenberg.
\newblock Quantum gas magnifier for sub-lattice-resolved imaging of 3d quantum
  systems.
\newblock {\em Nature}, 599(7886):571--575, 2021.

\bibitem{GiamarchiRuegg-12}
Thierry Giamarchi, Christian Rüegg, and Oleg Tchernyshyov.
\newblock Bose–einstein condensation in magnetic insulators.
\newblock {\em Nature Physics}, 4(3):198--204, 2008.

\bibitem{PhysRevB.93.100402}
Yaroslav Tserkovnyak, Scott~A. Bender, Rembert~A. Duine, and Benedetta Flebus.
\newblock Bose-einstein condensation of magnons pumped by the bulk spin seebeck
  effect.
\newblock {\em Phys. Rev. B}, 93:100402(R), Mar 2016.

\bibitem{Millar:2016cjp}
Alexander~J. Millar, Georg~G. Raffelt, Javier Redondo, and Frank~D. Steffen.
\newblock {Dielectric Haloscopes to Search for Axion Dark Matter: Theoretical
  Foundations}.
\newblock {\em JCAP}, 01:061, 2017.

\bibitem{2009PhRvL.103k1301S}
P.~{Sikivie} and Q.~{Yang}.
\newblock {Bose-Einstein Condensation of Dark Matter Axions}.
\newblock {\em \prl}, 103(11):111301, September 2009.

\bibitem{2010AIPC.1274...91Y}
Q.~{Yang}.
\newblock {Axion Bose-Einstein Condensation: a model beyond Cold Dark Matter}.
\newblock In David~B. {Tanner} and Karl~A. {van Bibber}, editors, {\em Axions
  2010}, volume 1274 of {\em American Institute of Physics Conference Series},
  pages 91--96. AIP, August 2010.

\bibitem{2010arXiv1012.1553S}
Pierre {Sikivie}.
\newblock {The dark matter is mostly an axion BEC}.
\newblock {\em arXiv e-prints}, page arXiv:1012.1553, December 2010.

\bibitem{2012PhRvL.108f1304E}
O.~{Erken}, P.~{Sikivie}, H.~{Tam}, and Q.~{Yang}.
\newblock {Axion Dark Matter and Cosmological Parameters}.
\newblock {\em \prl}, 108(6):061304, February 2012.

\bibitem{2011arXiv1111.3976E}
Ozgur {Erken}, Pierre {Sikivie}, Heywood {Tam}, and Qiaoli {Yang}.
\newblock {Axion BEC Dark Matter}.
\newblock {\em arXiv e-prints}, page arXiv:1111.3976, November 2011.

\bibitem{SantraBaals-13}
Bodhaditya Santra, Christian Baals, Ralf Labouvie, Aranya~B. Bhattacherjee,
  Axel Pelster, and Herwig Ott.
\newblock Measuring finite-range phase coherence in an optical lattice using
  talbot interferometry.
\newblock {\em Nature Communications}, 8(1):15601, 2017.

\bibitem{Koster_2026}
Malte Koster, Matthias~R. Schweizer, Timo Noack, Vitaliy~I. Vasyuchka,
  Dmytro~A. Bozhko, Burkard Hillebrands, Mathias Weiler, Alexander~A. Serga,
  and Georg von Freymann.
\newblock Emergence of phase coherence in a magnon bose–einstein condensate.
\newblock {\em Nature Physics}, 2026.

\bibitem{Jalali_Mola_2026}
Zahra Jalali-Mola, Niklas Käming, Luca Asteria, Utso Bhattacharya, Ravindra~W.
  Chhajlany, Klaus Sengstock, Maciej Lewenstein, Tobias Grass, and Christof
  Weitenberg.
\newblock Anomalous fluctuations of bose-einstein condensates in optical
  lattices.
\newblock {\em Physical Review Letters}, 136(8), February 2026.

\bibitem{SchneiderBracher-14}
Michael Schneider, Thomas Brächer, David Breitbach, Viktor Lauer, Philipp
  Pirro, Dmytro~A. Bozhko, Halyna~Yu. Musiienko-Shmarova, Björn Heinz,
  Qi~Wang, Thomas Meyer, Frank Heussner, Sascha Keller, Evangelos~Th.
  Papaioannou, Bert Lägel, Thomas Löber, Carsten Dubs, Andrei~N. Slavin,
  Vasyl~S. Tiberkevich, Alexander~A. Serga, Burkard Hillebrands, and Andrii~V.
  Chumak.
\newblock Bose–einstein condensation of quasiparticles by rapid cooling.
\newblock {\em Nature Nanotechnology}, 15(6):457--461, 2020.

\bibitem{MoritaYoshioka-15}
Yusuke Morita, Kosuke Yoshioka, and Makoto Kuwata-Gonokami.
\newblock Observation of bose-einstein condensates of excitons in a bulk
  semiconductor.
\newblock {\em Nature Communications}, 13(1):5388, 2022.

\bibitem{8wdy-2zbw}
J.~Khatua, S.~M. Kumawat, G.~Senthil Murugan, C.-L. Huang, Heung-Sik Kim,
  K.~Sritharan, R.~Sankar, and Kwang-Yong Choi.
\newblock Possible bose-einstein condensation of magnons in an
  $s$=$\frac{5}{2}$ honeycomb lattice.
\newblock {\em Phys. Rev. B}, 112:134422, Oct 2025.

\bibitem{Roising:2021lpv}
Henrik~S. R\o{}ising, Benjo Fraser, Sin\'ead~M. Griffin, Sumanta Bandyopadhyay,
  Aditi Mahabir, Sang-Wook Cheong, and Alexander~V. Balatsky.
\newblock Axion-matter coupling in multiferroics.
\newblock {\em Phys. Rev. Res.}, 3:033236, Sep 2021.

\bibitem{ArtyukhinDelaney-17}
Sergey Artyukhin, Kris~T. Delaney, Nicola~A. Spaldin, and Maxim Mostovoy.
\newblock Landau theory of topological defects in multiferroic hexagonal
  manganites.
\newblock {\em Nature Materials}, 13(1):42--49, 2014.

\bibitem{PhysRevMaterials.1.014401}
Dominik~M. Juraschek, Michael Fechner, Alexander~V. Balatsky, and Nicola~A.
  Spaldin.
\newblock Dynamical multiferroicity.
\newblock {\em Phys. Rev. Mater.}, 1:014401, Jun 2017.

\bibitem{Wilson2026LightIS}
Evan~M. Wilson, Hou-Tong Chen, and Alexander~V Balatsky.
\newblock Light induced superconducting diode effect in patterned films.
\newblock 2026.

\bibitem{Chen_2024}
Shuai~A. Chen and K.~T. Law.
\newblock Ginzburg-landau theory of flat-band superconductors with quantum
  metric.
\newblock {\em Physical Review Letters}, 132(2), January 2024.

\bibitem{xue2024case}
Cun Xue, Qing-Yu Wang, Han-Xi Ren, An~He, and A.~V. Silhanek.
\newblock Case studies on time-dependent ginzburg-landau simulations for
  superconducting applications, 2024.

\bibitem{Harbick_2025}
Aiden~V. Harbick and Mark~K. Transtrum.
\newblock Time-dependent ginzburg-landau framework for sample-specific
  simulation of superconductors for radio-frequency applications.
\newblock {\em Physical Review B}, 112(9), 2025.

\bibitem{PhysRevLett.97.167204}
M.~Cl\'emancey, H.~Mayaffre, C.~Berthier, M.~Horvati\ifmmode~\acute{c}\else
  \'{c}\fi{}, J.-B. Fouet, S.~Miyahara, F.~Mila, B.~Chiari, and O.~Piovesana.
\newblock Field-induced staggered magnetization and magnetic ordering in
  ${\mathrm{cu}}_{2}({\mathrm{c}}_{5}{\mathrm{h}}_{12}{\mathrm{n}}_{2}{)}_{2}{\mathrm{cl}}_{4}$.
\newblock {\em Phys. Rev. Lett.}, 97:167204, Oct 2006.

\bibitem{PhysRevB.94.064418}
F.~Qian, H.~Wilhelm, A.~Aqeel, T.~T.~M. Palstra, A.~J.~E. Lefering, E.~H.
  Br\"uck, and C.~Pappas.
\newblock Phase diagram and magnetic relaxation phenomena in
  ${\mathrm{cu}}_{2}{\mathrm{oseo}}_{3}$.
\newblock {\em Phys. Rev. B}, 94:064418, Aug 2016.

\bibitem{PhysRevB.85.224413}
M.~Belesi, I.~Rousochatzakis, M.~Abid, U.~K. R\"o\ss{}ler, H.~Berger, and
  J.-Ph. Ansermet.
\newblock Magnetoelectric effects in single crystals of the cubic ferrimagnetic
  helimagnet cu${}_{2}$oseo${}_{3}$.
\newblock {\em Phys. Rev. B}, 85:224413, Jun 2012.

\bibitem{OkamuraKagawa-10}
Y.~Okamura, F.~Kagawa, S.~Seki, and Y.~Tokura.
\newblock Transition to and from the skyrmion lattice phase by electric fields
  in a magnetoelectric compound.
\newblock {\em Nature Communications}, 7(1):12669, 2016.

\bibitem{PhysRevLett.113.107203}
J.~S. White, K.~Pr\ifmmode~\check{s}\else \v{s}\fi{}a, P.~Huang, A.~A. Omrani,
  I.~\ifmmode \check{Z}\else \v{Z}\fi{}ivkovi\ifmmode~\acute{c}\else
  \'{c}\fi{}, M.~Bartkowiak, H.~Berger, A.~Magrez, J.~L. Gavilano, G.~Nagy,
  J.~Zang, and H.~M. R\o{}nnow.
\newblock Electric-field-induced skyrmion distortion and giant lattice rotation
  in the magnetoelectric insulator ${\mathrm{cu}}_{2}{\mathrm{oseo}}_{3}$.
\newblock {\em Phys. Rev. Lett.}, 113:107203, Sep 2014.

\bibitem{s9l2-m9tt}
Yang Cao, Tong Li, Na~Lei, Liyang Liao, Baoshan Cui, Li~Xi, Dahai Wei, Tao Yu,
  Yoshichika Otani, Desheng Xue, and Dezheng Yang.
\newblock Inverse acoustic spin hall effect in heavy metal-ferromagnet
  bilayers.
\newblock {\em Phys. Rev. Lett.}, 135:246705, Dec 2025.

\bibitem{b6gt-w5xy}
T.~Nomura, I.~Rousochatzakis, O.~Janson, M.~Gen, X.-G. Zhou, Y.~Ishii, S.~Seki,
  Y.~Kohama, and Y.~H. Matsuda.
\newblock Quintuplet condensation in the skyrmionic insulator
  ${\mathrm{cu}}_{2}{\mathrm{oseo}}_{3}$ at ultrahigh magnetic fields.
\newblock {\em Phys. Rev. Lett.}, 136:076703, Feb 2026.

\bibitem{ManchonKoo-16}
A.~Manchon, H.~C. Koo, J.~Nitta, S.~M. Frolov, and R.~A. Duine.
\newblock New perspectives for rashba spin–orbit coupling.
\newblock {\em Nature Materials}, 14(9):871--882, 2015.

\bibitem{2014PhRvX...4c1045B}
Sumilan {Banerjee}, James {Rowland}, Onur {Erten}, and Mohit {Randeria}.
\newblock {Enhanced Stability of Skyrmions in Two-Dimensional Chiral Magnets
  with Rashba Spin-Orbit Coupling}.
\newblock {\em Physical Review X}, 4(3):031045, July 2014.

\bibitem{2025arXiv250417772L}
Paul {Leask} and Martin {Speight}.
\newblock {Demagnetization in micromagnetics: magnetostatic self-interactions
  of bulk chiral magnetic skyrmions}.
\newblock {\em arXiv e-prints}, page arXiv:2504.17772, April 2025.

\bibitem{2025arXiv250602192G}
Doried {Ghader} and Bilal {Jabakhanji}.
\newblock {Impact of the honeycomb spin-lattice on topological magnons and edge
  states in ferromagnetic 2D skyrmion crystals}.
\newblock {\em arXiv e-prints}, page arXiv:2506.02192, June 2025.

\bibitem{2015NatMa..14..871M}
A.~{Manchon}, H.~C. {Koo}, J.~{Nitta}, S.~M. {Frolov}, and R.~A. {Duine}.
\newblock {New perspectives for Rashba spin{\textendash}orbit coupling}.
\newblock {\em Nature Materials}, 14(9):871--882, September 2015.

\bibitem{2024arXiv241201631Y}
F{\i}rat {Y{\i}lmaz}.
\newblock {The dynamical enhancement of Dzyaloshinskii-Moriya interaction in
  lattice Anderson impurity model}.
\newblock {\em arXiv e-prints}, page arXiv:2412.01631, December 2024.

\bibitem{10.1143/PTP.63.387}
Shinobu Hikami and Toshihiko Tsuneto.
\newblock Phase transition of quasi-two dimensional planar system.
\newblock {\em Progress of Theoretical Physics}, 63(2):387--401, 02 1980.

\bibitem{DuZhang-9}
K.~Du, M.~Zhang, C.~Dai, Z.~N. Zhou, Y.~W. Xie, Z.~H. Ren, H.~Tian, L.~Q. Chen,
  Gustaaf Van~Tendeloo, and Z.~Zhang.
\newblock Manipulating topological transformations of polar structures through
  real-time observation of the dynamic polarization evolution.
\newblock {\em Nature Communications}, 10(1):4864, 2019.

\bibitem{PhysRevX.4.031045}
Sumilan Banerjee, James Rowland, Onur Erten, and Mohit Randeria.
\newblock Enhanced stability of skyrmions in two-dimensional chiral magnets
  with rashba spin-orbit coupling.
\newblock {\em Phys. Rev. X}, 4:031045, Sep 2014.

\end{thebibliography}

\newpage
\appendix

\section{Tables of Data and Parameters}
\begin{table}[htbp]
	\centering
	\caption{Experimental Measurement Data} 
	\label{tab:emd} 
	\begin{tabular}{l c c c c c c c } 
		\toprule
		No. & Parameter Name & Symbol &  I$_{1.5}$/T$_3$ & I$_3$/T$_6$ & I$_3$/B$_6$ & I$_3$/B$_8$ & Unit\\ 
		\midrule
		1 & ME phase transition temperature & $T_c$ &  46 & 77 &166& 203 & K \\ 
		2 & critical polarization  & $P_0$ & 0.02& 0.2& 0.55& 0.8 &$\mu\text{C}/\text{cm}^2$ \\ 
		3 & critical magnetization & $M_0$ & 2.2&2.7& 3.5& 4.2&~emu/cm$^3$ \\ 
		4 & static AQ angle & $\theta_0$& 0.194& 0.423& 0.397&0.701 & / \\ 
		5 & frequency shift & $\Delta f$ &/& 0.19&0.42&/ & GHz \\ 
		\bottomrule
	\end{tabular}
\end{table}

\begin{table}[htbp]
	\centering
	\caption{Fitting Parameters} 
	\label{tab:fp} 
	\begin{tabular}{l c c c c c c c } 
		\toprule
		No. & Parameter Name & Symbol & I$_{1.5}$/T$_3$&I$_3$/T$_6$&I$_3$/B$_6$&I$_3$/B$_8$& Unit\\ 
		\midrule
		1 & static effective AQ mass & $m_a$&0.08&0.39&0.03&0.16&meV \\ 
		2 & phenomenological coefficient &$\lambda_M$&1.5&22& 0.08& 1.6&kOe$^{-2}$ \\ 
		3 & phenomenological coefficient & $\lambda_\theta$& 0.005& 0.0315& 2& 0.3&kOe$^2$ \\ 
		4 & static AQ angle & $(\theta_0)_\text{max}$& 0.196& 0.410& 0.395& 0.694&/ \\ 
		5 &  frequency shift & $\Delta f$ & 0.114& /&/& 0.514&GHz \\ 
		6 &  wave vector& $k$&16.2&18.5&23.8&25.4&rad$\cdot\mu$m$^{-1}$ \\ 
		7 & interfacial DMI &$D_{\text{int}}$& 0.13& 0.18& 0.32&0.36&mJ/m$^2$ \\ 
		\bottomrule
	\end{tabular}
\end{table}

\begin{table}[htbp]
	\centering
	\caption{Cited Parameters} 
	\label{tab:cp} 
	\begin{tabular}{l c c c c } 
		\toprule
		No. & Parameter Name & Symbol & I$_{1.5}$/T$_3$& Unit\\ 
		\midrule
		1 & phenomenological coefficient & $\alpha_M$&1&meV$^{-2}$ \\ 
		2 & phenomenological coefficient  &  $\alpha_P$&$10^{-2}$&$\text{meV}^{-2}$ \\ 
		3 & damping coefficient &$\eta_M$&5&$\mu\text{eV}$\\ 
		4 & damping coefficient & $\eta_P$&5&$\mu\text{eV}$ \\ 
		5 & effective mass & $m_P$&$4.5\times10^{-2}$ &/ \\ 
		6 & phenomenological coefficient & $\gamma_M$ &$10^{-5}$&$\text{K}^{-1}$ \\ 
		7 & AQ-photon coupling & $g_{a\gamma}$ &0.225&$\text{meV}^{-1}$ \\ 
		8 & phenomenological coefficient& $\beta$ &1&meV \\ 
		9 & phenomenological coefficient & $\delta$ &0.05&/\\ 
		10 &  gyromagnetic ratio &$\gamma$&$1.76\times10^{11}$& rad$\cdot$s$^{-1}\cdot$T$^{-1}$ \\ 
		11 & saturation magnetization& $M_s$&$2\times10^6$&A/m \\ 
		\bottomrule
	\end{tabular}
\end{table}

\section{The extended GL theory}

The classical equations of motion for M, P and $\theta$ are derived by 
\begin{align}
	\frac{\partial \mathcal{F}[\mathbf{M}, \mathbf{P},\theta]}{\partial M}&=\alpha_M\partial^2_t\mathbf{M}-\beta_M\nabla^2\mathbf{M}+\gamma_M(T-T_{C_M})\mathbf{M}+2\lambda_M|\mathbf{M}|^2\mathbf{M}+\frac{e^2}{8\pi^2}\theta\mathbf{P}=0,
	\\
	\frac{\partial \mathcal{F}[\mathbf{M}, \mathbf{P},\theta]}{\partial P}&=\alpha_P\partial^2_t\mathbf{P}-\beta_P\nabla^2\mathbf{P}+\gamma_P(T-T_{C_M})\mathbf{P}+2\lambda_P|\mathbf{P}|^2\mathbf{P}+\frac{e^2}{8\pi^2}\theta\mathbf{M}=0,
	\\
	\frac{\partial \mathcal{F}[\mathbf{M}, \mathbf{P},\theta]}{\partial \theta}&=f^2\partial_\mu\partial^\mu\theta+f^2m_a^2\theta+4\lambda_\theta|\theta|^2\theta+\frac{e^2}{4\pi^2}\mathbf{P}\cdot\mathbf{M}=0.
\end{align}
We consider spatially homogeneous solutions here, so that we have $\mathbf{P}(t;x)=P(t)\hat{e}, \mathbf{M}(t;x)=M(t)\hat{e}.$
We also have $|\partial_\mu\theta|^2=|\partial_t\theta|^2$    and  $\partial_\mu\partial^\mu\theta=\partial_t\partial^t\theta=\ddot{\theta}$, thus we can derive
\begin{align}
	\label{eq:MthetaP}	\alpha_M \ddot M +\gamma_M(T-T_{C_M}) M + 2 \lambda_M M^3 +\frac{e^2}{8\pi^2}\theta P\, &= \, 0, \\
	\label{eq:PthetaM}	\alpha_P \ddot P+\gamma_P(T-T_{C_P}) P + 2 \lambda_P P^3 +\frac{e^2}{8\pi^2}\theta M\, &= \, 0,\\
	\label{eq:thetaPM}	f^2\ddot{\theta}+f^2m_a^2\theta+4\lambda_\theta\theta^3+\frac{e^2}{4\pi^2}PM&=0.
\end{align}

The $\lambda_MM^3$ term in Eq. \ref{eq:MthetaP} is also called the cubic Gross-Pitaevskii (GP) term. The cubic GP term, which breaks time-reversal symmetry $\mathcal{T}$, is often used to describe BEC behavior, related to GP equations \cite{Jalali_Mola_2026}. Because of this deep connection between BEC and the ME phase transition, we naturally think that the BEC behavior of AQ may give a phenomenologically effective description for the ME phase transition. Thus, it is consistent with our treatment of AQs as bosons and the ME phase transition as a BEC process in the former section.

We set $M=M_0+\delta M, P=P_0+\delta P$ and $\theta=\theta_0+\delta\theta$, where $M_0$ and $P_0$ are the values of the static polarization and magnetization when the temperature equals the phase transition temperature, and $\theta_0$ is the static AQ angle.
We can get
\begin{align}
	\label{Eq:deltaM}\alpha_M\delta\ddot{M}+\alpha_M\eta_M \delta\dot{M}+m_M^2 \delta M+\frac{e^2}{8\pi^2}(\theta_0\delta P+P_0\delta\theta)&=0,\\
	\label{Eq:deltaP}\alpha_P\delta\ddot{P}+\alpha_P\eta_P \delta\dot{P}+m_P^2 \delta P+\frac{e^2}{8\pi^2}(\theta_0\delta M+M_0\delta\theta)&=0,\\
	\label{Eq:deltatheta}f^2\ddot{\delta\theta}+f^2\eta_\theta\dot{\delta\theta}+f^2m_a^2\delta\theta+12\lambda_\theta\theta_0^2\delta\theta+\frac{e^2}{4\pi^2}(P_0\delta M+\delta PM_0)&=0.
\end{align}
Here, we insert phenomenological damping terms  $\alpha_M\eta_M \delta\dot{M}$, $\alpha_P\eta_P \delta\dot{P}$ and $f^2\eta_\theta\dot{\delta\theta}$  by hand. Also, terms $\delta\theta\delta P$, $\delta\theta\delta M$ and $\delta P\delta M$ have been neglected.
We define the effective mass $
m_M^2 \equiv  \gamma_M(T-T_{C_M}) + 6 \lambda_M M_0^2$ and $m_P^2 \equiv \gamma_P(T-T_{C_P}) + 6 \lambda_P P_0^2$.
The solutions to the classical equations of motion for $\delta P$, $\delta M$ and $\delta \theta$ is expected to oscillate, $\delta\theta(t)=\delta\theta~e^{-i\omega_at}$,
$\delta M(t)=\delta M\ e^{-i\omega_at}$, $\delta P(t)=\delta P\ e^{-i\omega_at}$.
By bringing the above equations into Eq. \ref{Eq:deltatheta}, we get
\begin{align}
	-\omega_a^2\alpha_M\delta M-i\omega_a\alpha_M\eta_M\delta M+m_M^2\delta M+\frac{e^2}{8\pi^2}(\theta_0\delta P+P_0\delta\theta)&=0,\\
	-\omega_a^2\alpha_P\delta P-i\omega_a\alpha_P\eta_P\delta P+m_P^2\delta P+\frac{e^2}{8\pi^2}(\theta_0\delta M+M_0\delta\theta)&=0,\\
	-\omega_a^2f^2\delta\theta +f^2m_a^2\delta\theta -i\omega_af^2\eta_\theta\delta\theta +12\lambda_\theta\theta_0^2\delta\theta +\frac{e^2}{4\pi^2}(P_0\delta M+\delta PM_0)&=0.
\end{align}

Nonrelativistically, we set the AQ mass $m_a=\omega_a$. Through simplification, we can derive
\begin{align}
	\label{deltaM}
	\frac{\delta M}{\delta\theta}&=e^2\frac{e^2\theta_0P_0-8\pi^2M_0F_M(\omega_a,T)}{-e^4\theta_0^2+64\pi^2F_M(\omega_a,T)F_P(\omega_a)},\\
	\label{deltaP}
	\frac{\delta P}{\delta\theta}&=e^2\frac{e^2\theta_0M_0-8\pi^2P_0F_P(\omega_a)}{-e^4\theta_0^2+64\pi^2F_M(\omega_a,T)F_P(\omega_a)},\\
	\label{eq:deltatheta}
	\delta\theta&=\frac{e^2}{4\pi^2}\frac{P_0\delta M+\delta PM_0}{f^2(m_a^2-\omega_a^2)+12\theta_0^2\lambda_\theta-i\omega_af^2\eta_\theta},
\end{align}
where we define functions $F_M(\omega_a,T)\equiv i \alpha_M\eta_M \omega_a-\alpha_M\omega_a^2+m_M^2$ and $F_P(\omega_a)\equiv i\alpha_P \eta_P \omega_a-\alpha_P\omega_a^2+m_P^2$. Here, $
m_M^2 \equiv  \gamma_M(T-T_{C_M}) + 6 \lambda_M M_0^2$ and $m_P^2 \equiv \gamma_P(T-T_{C_P}) + 6 \lambda_P P_0^2 $ are the effective masses about equilibrium metioned above. 
Putting Eqs. \ref{deltaM} and \ref{deltaP} into Eq. \ref{eq:deltatheta}, we can get the relation between the parameters in the above equations

\begin{align}\label{Eq:theta0wa3}
	{12\lambda_\theta}\theta_0^2+\frac{e^4}{2\pi^4}\frac{4\pi^2(P_0^2F_P(\omega_a)+M_0^2F_M(\omega_a,T))-e^2\theta_0P_0M_0}{e^4\theta_0^2-64\pi^2F_M(\omega_a,T)F_P(\omega_a)}-i\omega_af^2\eta_\theta=0.
\end{align}

Because we have $T=T_c=T_{C_M}=T_{C_P}$, so the terms $T-T_{C_M}$ and $T-T_{C_P}$ in the definition of $m_M$ and $m_P$ equal zero. Thus Eq. \ref{Eq:theta0wa3} becomes
\begin{align}\label{Eq:theta0wa4}
	{12\lambda_\theta}\theta_0^2+\frac{e^4}{2\pi^4}\frac{4\pi^2(P_0^2F_P(\omega_a)+M_0^2F_M(\omega_a))-e^2\theta_0P_0M_0}{e^4\theta_0^2-64\pi^2F_M(\omega_a)F_P(\omega_a)}-i\omega_af^2\eta_\theta=0.
\end{align}
Eq. \ref{Eq:theta0wa4} does not relate to the temperature or phase transition temperature, and we can solve $\theta_0$ as a function of the frequency of AQ $\omega_a$, as shown in the right panel of Fig. \ref{fig:thetaT}. Eq. \ref{Eq:theta0wa4} is a quartic equation of $\theta_0$ with respect to $\omega_a$ as follows
\begin{align}\label{Eq:theta0wa5}
	12e^4\lambda_\theta\theta_0^4-768\pi^2\lambda_\theta\theta_0^2F_M(\omega_a)F_P(\omega_a)-i\omega_af^2\eta_\theta e^4\theta_0^2-\frac{e^6}{2\pi^4}\theta_0P_0M_0\\\notag
	+\frac{2e^4}{\pi^2}(P_0^2F_P(\omega_a)+M_0^2F_M(\omega_a))+64i\omega_af^2\eta_\theta\pi^2F_M(\omega_a)F_P(\omega_a)=0.
\end{align}
For the plots, we take the physically meaningful solutions to the quartic equation.

\clearpage

\end{document}